\documentclass[12pt]{article}
\usepackage{amsmath}
\usepackage{times}
\usepackage{graphicx}
\graphicspath{{figures/}{paper/figures/}}
\usepackage{color}
\usepackage{multirow}
\usepackage{float}
\usepackage{amsfonts}

\usepackage[authoryear]{natbib}
\usepackage{caption}
\usepackage{rotating}
\usepackage{bbm}
\usepackage{latexsym}
\usepackage{tikz}
\usetikzlibrary{arrows.meta, positioning, calc, decorations.pathreplacing}
\usepackage[normalem]{ulem}
\usepackage{xcolor}
\definecolor{OliveGreen}{cmyk}{0.64,0,0.95,0.40}
\definecolor{DarkGreen}{RGB}{1,50,32}

\newcommand{\captionfonts}{\normalsize}

\makeatletter  
\long\def\@makecaption#1#2{%
  \vskip\abovecaptionskip
  \sbox\@tempboxa{{\captionfonts #1: #2}}%
  \ifdim \wd\@tempboxa >\hsize
    {\captionfonts #1: #2\par}
  \else
    \hbox to\hsize{\hfil\box\@tempboxa\hfil}%
  \fi
  \vskip\belowcaptionskip}
\makeatother   
\begin{document}
\hspace{13.9cm}1

\ \vspace{20mm}\\

{\LARGE Geometric Reliability of Neural Population Codes: Sampling Calibration and Within-Session Nonstationarity}

\ \\
{\bf \large Prashant C. Raju}\\
%

%\ \\[-2mm]

\thispagestyle{empty}
\markboth{}{NC instructions}
\ \vspace{-0mm}\\

\begin{center} {\bf Abstract} \end{center}
Trial-to-trial variability limits how reliably the representational geometry of a neural population can be estimated, but comparisons across populations are sensitive to neuron count, trial count, response quality, and clustered sampling. We quantified within-session geometric reliability using Shesha, an implementation of the established split-data RSA reliability statistic---the Spearman correlation between representational dissimilarity matrices (RDMs) estimated from independent trial subsets. The analysis included all 39 sessions in the Steinmetz Neuropixels release and olfactory bulb and piriform cortex recordings from Bolding and Franks. For the Steinmetz data, we matched neurons and repetitions across area-by-session recordings, compared observed reliability with a stationary residual-bootstrap expectation, and used mouse-level or mouse-clustered inference. Mean matched reliability was 0.0402 across 312 area-by-session recordings. Descriptive regional differences in reliability were present, but neither the quality-adjusted regional omnibus nor reliability in excess of the stationary benchmark survived correction. Evidence was strongest for temporal effects: interleaving the same early and late trial pools increased RDM reliability relative to keeping them blocked ($\Delta=0.02666$, $q=0.001953$), and RDM similarity declined with within-session block lag (mean mouse-level slope $=-0.01912$, $q=0.001953$; $n=10$ mice). In the designated outer-cross-fitted analyses, reliability was not associated with choice-direction coupling, stimulus decoding, or response-direction decoding after correction. The olfactory comparisons were descriptive because only five session pairs comparing PCx with OB and three comparing Control with TeLC were available, and animal identities were unknown. In held-out simulations, associative recurrence outperformed feedforward subspace denoising but not divisive normalization. Representational geometry became less reproducible with temporal separation within a session. Comparisons of geometric reliability across neural populations therefore require sampling calibration and independent inference.

%%%%%%%%%%%

\section{Introduction}
\label{sec1}

Repeated presentations of the same stimulus do not evoke identical neural responses. Spike counts vary from trial to trial even under nominally fixed conditions, and this variability depends on firing rate, shared input, and fluctuations in neuronal gain~\citep{Schiller1976, Shadlen1998, Goris2014}. At the population level, this variability limits the precision of condition means, coding axes, and pairwise relationships estimated from finite samples. We study the corresponding multivariate reliability problem: how reproducibly can a population's representational geometry be observed across independent trials within one recording session?

Longitudinal representational drift concerns changes in neural representations across days or weeks, often despite stable behavior~\citep{Ziv2013, Driscoll2017, rule2019causes, Rule2020}. Work on drift has examined the preservation of task-relevant information, structured motion through representational space, and compensatory readout mechanisms~\citep{Aitken2022, Rule2022, Qin2023, Driscoll2022}. Our analysis concerns a shorter timescale. It asks how reliably geometry can be estimated from repeated trials within a session and whether that reliability changes with temporal separation on the scale of minutes. We call these quantities \emph{within-session geometric reliability} and \emph{within-session nonstationarity}; neither measures drift across days.

Representational similarity analysis~\citep[RSA;][]{Kriegeskorte2008} uses a representational dissimilarity matrix (RDM) to summarize all pairwise distances among condition-averaged population responses. Correlating RDMs estimated from independent data splits is an established RSA reliability or noise-ceiling procedure~\citep{Nili2014, Walther2016, diedrichsen2017representational}. We use the Spearman correlation between two split-half RDM vectors, implemented as Shesha~\citep{raju2026geometric,shesha2026}, and calibrate it across neural populations, recording areas, and circuit hypotheses.

Cross-area comparisons of split-half reliability depend on several sampling and response properties. The estimate can change with the number of recorded units, repetitions per condition, firing rate, stimulus-driven variance, and residual noise. Area-by-session observations are crossed rather than independent because the same session contributes several areas and the same area is sampled in several sessions. We therefore match unit and trial counts, compare observed reliability with a stationary residual-bootstrap expectation, quantify firing rate, count-window Fano factor, and signal-to-noise ratio, and conduct inference at the mouse or appropriately clustered level. The resulting tests ask whether differences exceed those expected from sampling and measured response quality.

Decoding and RDM reliability measure different properties. A decoder tests whether a specified variable is recoverable from population activity~\citep{Hebart2018}. Split-half RDM correlation tests whether the full pattern of pairwise condition distances is reproducible. Performance on one measure need not predict performance on the other. Shared trial noise can induce correlations among reliability, decoding, and behavior when all three are estimated from the same data. We use outer cross-fitting to separate them: one set of trials estimates geometric reliability, and disjoint trials estimate choice-direction coupling and decoder performance. The association models also adjust for measured quality covariates.

Recurrent associative dynamics can complete stimulus patterns from degraded input~\citep{Hopfield1982, Haberly2001}, and recurrent circuitry is important for stable odor representations in piriform cortex~\citep{Bolding2018, Bolding2020}. However, feedforward denoising and divisive normalization can also stabilize population geometry. The recordings from Bolding and Franks provide descriptive circuit context, while held-out simulations compare associative recurrence with matched nonrecurrent alternatives. This design tests recurrent stabilization against mechanisms that can produce similar gains in reliability.

We apply this framework to all 39 sessions in the \cite{steinmetz2019distributed} Neuropixels release and to the PCX-1 olfactory recordings~\citep{Bolding2018}. The analyses ask four questions: whether sampling-calibrated geometric reliability varies among regions after quality adjustment; whether geometry becomes less reproducible with greater within-session temporal separation; whether reliability predicts choice-direction coupling or decoding under outer cross-fitting; and whether associative recurrence outperforms matched alternative stabilization mechanisms. Within-session temporal nonstationarity received the strongest empirical support. The regional and functional effects are uncertain, the olfactory comparisons are descriptive, and the model comparison does not identify a unique circuit mechanism.

\section{Results}
\label{sec2}

\subsection{Sampling-calibrated Steinmetz reliability}

\begin{figure}[H]
  \centering
  \includegraphics[width=\linewidth]{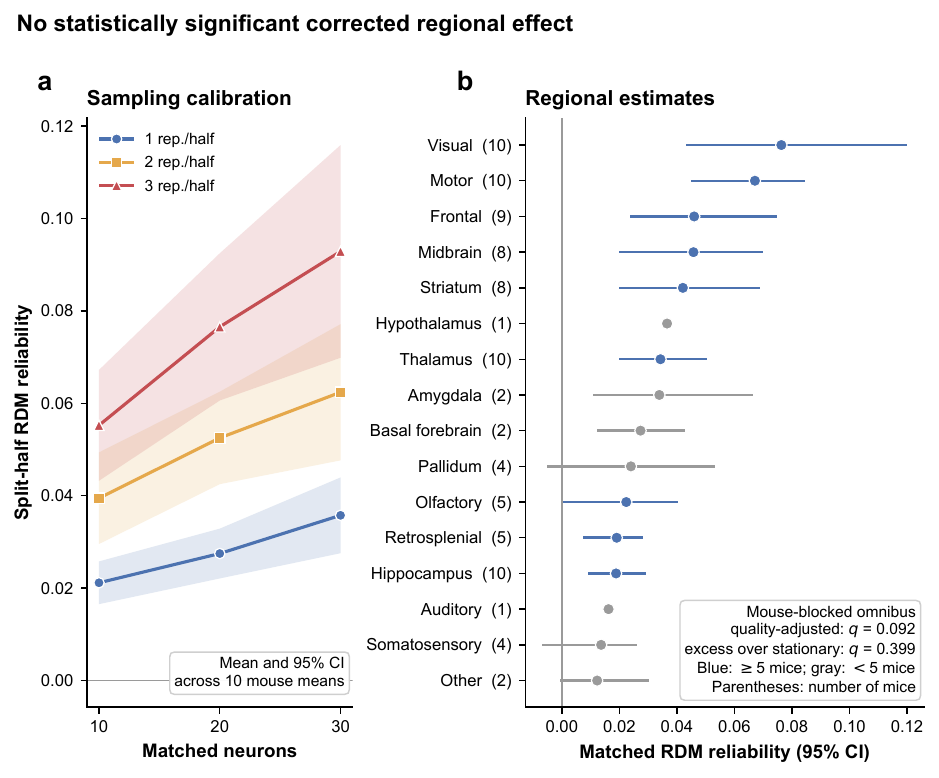}
  \caption{%
    \textbf{Sampling-calibrated split-half reliability and regional
estimates.} (a) Mean split-half RDM rank reliability across mouse-level
averages as a function of matched neuron count and repetitions per condition
per half. Shading denotes 95\% \textit{t} intervals across mice. (b) Regional means
and hierarchical 95\% bootstrap intervals at the primary specification of 10
neurons and two repetitions per condition per half. Parentheses give the
number of mice contributing to each region; sparsely sampled regions with
fewer than five mice are gray. Neither the quality-adjusted regional omnibus
nor the excess-over-stationary omnibus survived correction across regional
endpoints.
  }
  \label{fig:sampling}
\end{figure}
 
The sampling-calibrated analysis of the Steinmetz recordings~\citep{steinmetz2019distributed} included 312 area-sessions from 39 sessions and 10 mice at the primary specification of 10 neurons and two repetitions per condition in each half. Mean matched split-half RDM reliability was 0.0402, compared with a mean stationary residual-bootstrap expectation of 0.1095, yielding a descriptive mean excess of $-0.0693$. The full calibration grid contained 2,598 area-session-by-design rows. Each recording-level estimate averaged up to 100 neuron/trial subsampling draws and 100 stationary-bootstrap draws; these Monte Carlo draws quantified uncertainty within a recording and were not treated as biological observations. The bootstrap independently resampled whole residual vectors within condition and therefore removed trial order and serial dependence. The negative excess means that observed reliability was lower than expected under this fitted exchangeable-residual model; it is not a null ceiling or an estimate of a specific temporal process.

\subsection{Regional inference}

Regional inference reduced the primary sample to 91 mouse-by-region observations spanning 16 coarse regions and 10 mice. The unadjusted regional omnibus was $F_{15,66}=1.805$, $p=0.0364$, $q=0.0918$. After adjustment for SNR, mean firing rate, and count-window Fano factor, the matched-reliability omnibus was $F_{15,63}=1.771$, $p=0.0459$, $q=0.0918$. Regional variation was not detected for excess over the stationary calibration ($F_{15,66}=1.203$, $p=0.2995$, $q=0.3993$) or for quality-adjusted crossnobis distance strength ($F_{15,63}=1.021$, $p=0.4316$, $q=0.4316$).

Descriptively, matched reliability was 0.0763 (95\% hierarchical bootstrap CI $[0.0438,0.1196]$; 37 area-sessions, 25 sessions, 10 mice) in visual areas, 0.0671 ($[0.0455,0.0840]$; 19 area-sessions, 19 sessions, 10 mice) in motor areas, and 0.0189 ($[0.0096,0.0287]$; 58 area-sessions, 26 sessions, 10 mice) in hippocampal areas. Same-session contrasts were positive for visual minus hippocampal areas ($\Delta=0.0654$; 20 session pairs reduced to 10 mouse-level contrasts; $p=0.003906$, $q=0.1797$) and visual minus thalamic areas ($\Delta=0.0666$; 15 session pairs reduced to 10 mouse-level contrasts; $p=0.003906$, $q=0.1797$), but neither survived correction across pairwise comparisons. The data therefore show descriptive regional differences but do not establish a corrected regional effect or pairwise ordering.

\subsection{Within-session temporal geometry}

\begin{figure}[H]
  \centering
  \includegraphics[width=\linewidth]{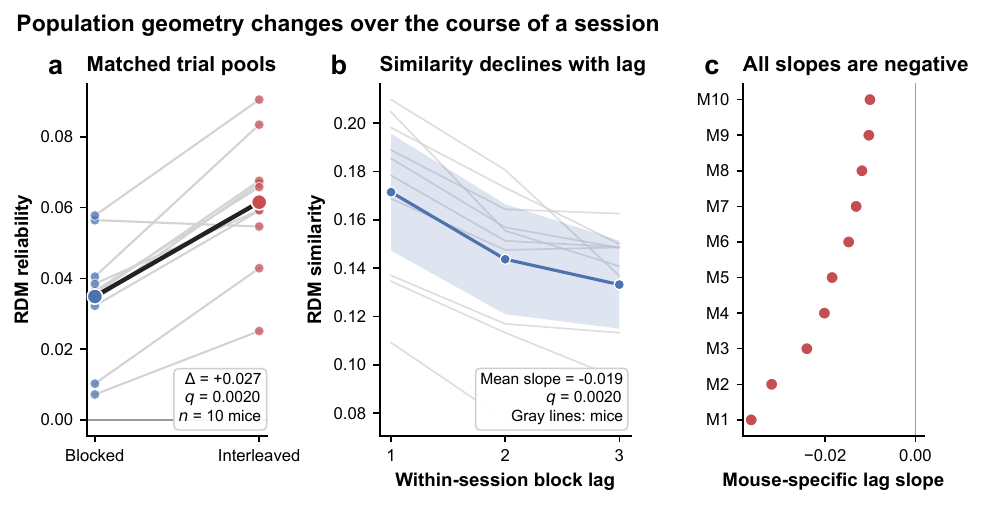}
  \caption{%
    \textbf{Representational geometry becomes less reproducible with
within-session temporal separation.} (a) Mouse-level geometric reliability
for temporally blocked and interleaved condition means constructed from the
same trials, neuron count, and repetitions. Thin lines denote mice and the
heavy line denotes the across-mouse mean. (b) RDM similarity by separation
between successive within-session repetition blocks. Gray lines denote mice;
the colored line and shading denote the across-mouse mean and 95% *t*
interval. (c) Mouse-specific slopes of RDM similarity against block lag. The
interleaved-minus-blocked difference and lag slope survived correction across
the two confirmatory temporal hypotheses. These analyses measure
within-session nonstationarity, not change across days.
  }
  \label{fig:temporal}
\end{figure}

The clearest positive result concerned temporal separation within sessions. Matched blocked and interleaved estimates were available for 311 area-sessions from 39 sessions and 10 mice. Interleaving the same early and late condition-wise trial pools increased RDM similarity relative to keeping them temporally blocked (mean mouse-level difference $=0.02666$, $n=10$ mice, one-sided $p=0.001953$, $q=0.001953$). In the complementary four-block analysis, RDM similarity decreased with block separation (mean within-mouse slope $=-0.01912$, $n=10$ mice, one-sided $p=0.0009766$, $q=0.001953$). The temporal decrement did not vary significantly among coarse regions ($F_{15,63}=1.607$, $p=0.0952$; 91 mouse-by-region observations, 16 regions, 10 mice). These endpoints provide evidence for within-session temporal nonstationarity of representational geometry; they do not address longitudinal change across days.

\subsection{Outer-cross-fitted choice and decoder null results}

\begin{figure}[H]
  \centering
  \includegraphics[width=\linewidth]{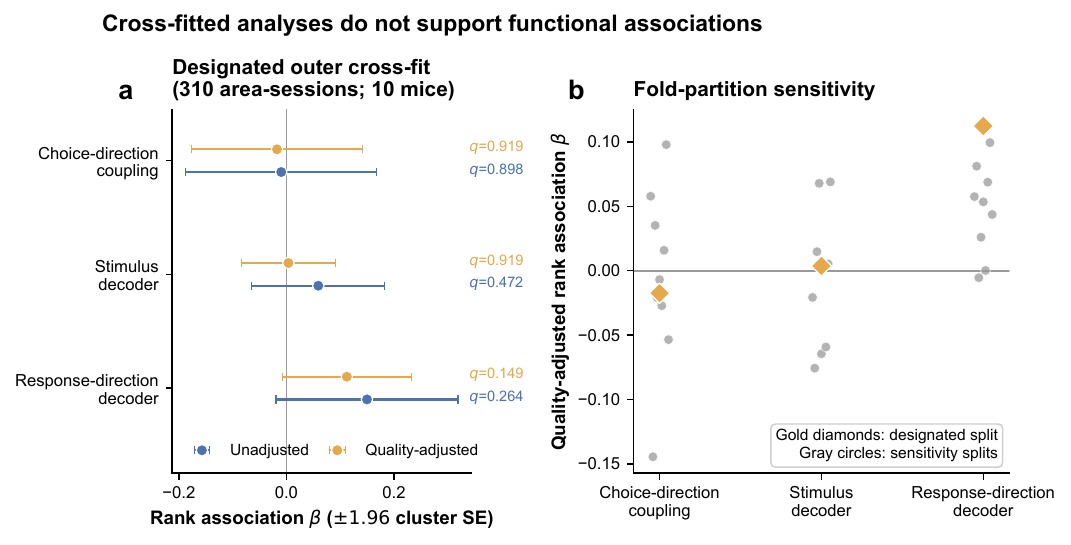}
  \caption{%
    \textbf{Outer-cross-fitted functional associations.} (a) Rank-regression
coefficients for the designated split, in which folds 0--1 estimated
geometric reliability and disjoint folds 2--3 estimated choice-direction
coupling and decoder outcomes. Error bars show 1.96 mouse-clustered standard
errors; displayed \textit{q} values are randomization-test \textit{p} values adjusted within
the unadjusted and quality-adjusted outcome families. (b) Quality-adjusted
coefficients across reciprocal and alternative fold partitions. Gold diamonds
mark the designated split and gray points mark sensitivity splits. No
association was supported in the designated confirmatory analysis.
  }
  \label{fig:outer_crossfit}
\end{figure}

The designated outer-cross-fit used folds 0--1 to estimate reliability and disjoint folds 2--3 to estimate stimulus-residualized choice-direction coupling and decoder performance. Of 313 candidate area-sessions from 39 sessions and 10 mice, each primary regression included 310 complete area-sessions spanning 72 fine areas. In this designated analysis, reliability was not associated with choice-direction coupling ($\beta=-0.0096$, $p=0.8984$, $q=0.8984$), stimulus decoding ($\beta=0.0593$, $p=0.3145$, $q=0.4717$), or response-direction decoding ($\beta=0.1498$, $p=0.08789$, $q=0.2637$). After adjustment for SNR, firing rate, Fano factor, and recorded neuron count, the corresponding estimates remained non-significant: choice-direction coupling, $\beta=-0.0175$, $p=0.8066$, $q=0.9189$; stimulus decoding, $\beta=0.0037$, $p=0.9189$, $q=0.9189$; and response-direction decoding, $\beta=0.1123$, $p=0.04980$, $q=0.1494$. Some non-designated fold partitions produced nominal unadjusted decoder associations, but these were not consistent after quality adjustment and were treated only as sensitivity analyses. Ten-neuron subsampling draws were averaged within area-session; these Monte Carlo draws, reciprocal fold assignments, and additional partitions did not increase the inferential sample size.

\subsection{Descriptive concentration generalization in the Bolding recordings}

\begin{figure}[H]
  \centering
  \includegraphics[width=\linewidth]{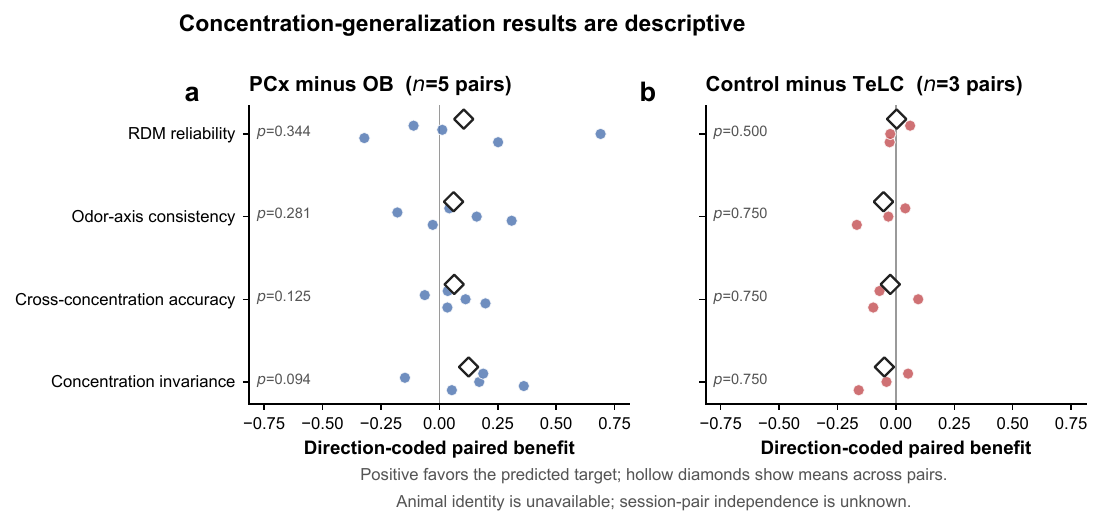}
  \caption{%
    \textbf{Descriptive concentration generalization in the Bolding--Franks
recordings.} Direction-coded paired benefits for five simultaneous PCx--OB
pairs (a) and three Control--TeLC pairs (b). Positive values favor the
specified target for every metric; black diamonds mark session-pair means.
Animal identifiers were unavailable, so independence among session pairs is
unknown and all displayed tests are descriptive rather than confirmatory.
Unit-subsampling draws were integrated within recordings and were not treated
as biological replicates.
  }
  \label{fig:bolding}
\end{figure}

The documented loading-A panel in the Bolding and Franks dataset~\citep{Bolding2018} yielded 20 recordings and eight complete, unit-matched session pairs: five simultaneous PCx--OB pairs and three Control--TeLC pairs. For PCx relative to OB, the direction-coded differences favored PCx for split-half reliability ($\Delta=0.104$, $p=0.3438$), odor-axis consistency across concentrations ($\Delta=0.061$, $p=0.2812$), leave-one-concentration-out odor decoding ($\Delta=0.063$, $p=0.125$), and a lower same-odor concentration-distance ratio ($\Delta=0.125$, $p=0.09375$; all $n=5$ session pairs). Control relative to TeLC showed little split-half difference ($\Delta=0.003$, $p=0.500$) and effects opposite the specified direction for odor-axis consistency ($\Delta=-0.053$, $p=0.750$), cross-concentration decoding ($\Delta=-0.024$, $p=0.750$), and the concentration-distance benefit ($\Delta=-0.049$, $p=0.750$; all $n=3$ session pairs). Animal identities were unavailable in the catalogs, so independence among session pairs is unknown; these sign-flip results are descriptive, no confirmatory $q$-values were assigned, and the 200 unit-subsampling draws per recording were not biological replicates.

\subsection{Qualified comparison of competing mechanisms}

\begin{figure}[p] % 'p' encourages it to take its own page
  \centering
  \includegraphics[width=\linewidth]{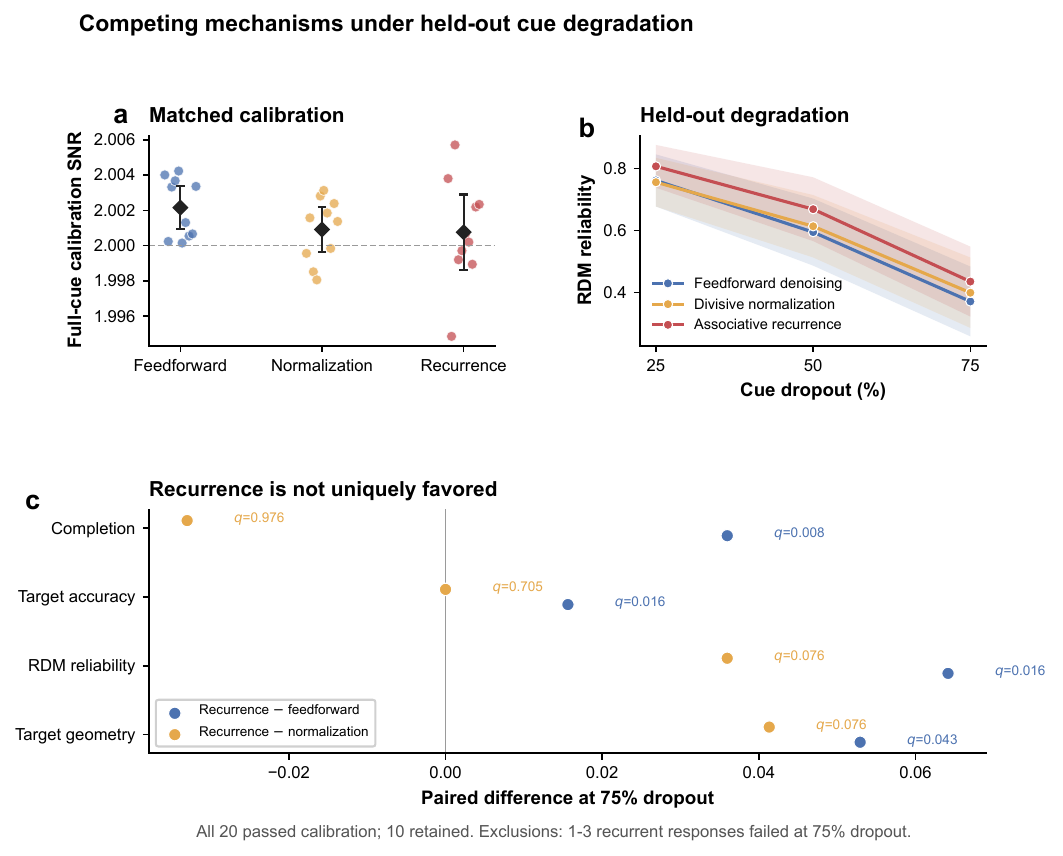}
  \caption{\textbf{Held-out comparison of competing computational mechanisms.}}
  \label{fig:mechanism}
\end{figure}

\begin{figure}[t] % 't' puts it at the top of the next page
  \ContinuedFloat
  \caption*{\textbf{Figure \ref{fig:mechanism} (continued):} 
    \textbf{Held-out comparison of competing computational mechanisms.} (a) Full-cue SNR after mechanism-specific calibration. Points denote eligible
networks and black diamonds with bars denote means and 95\% \textit{t} intervals.
(b) Split-half RDM reliability under held-out cue dropout for feedforward
subspace denoising, divisive normalization, and associative recurrence.
Shading denotes 95\% \textit{t} intervals across networks. (c) Paired differences
between recurrence and each control at 75\% dropout, with \textit{q} values adjusted
across eight comparisons. Recurrence exceeded feedforward denoising but not
divisive normalization. All 20 networks passed full-cue SNR calibration.
Ten were retained after held-out convergence screening; each excluded network
had one to three associative-recurrence responses fail the fixed-point
tolerance at 75\% dropout, and neither feedforward branch failed evaluation.
Inference is conditional on the retained networks.
  }
\end{figure}

As a supplementary model comparison, associative recurrence, feedforward subspace denoising, and divisive normalization were calibrated using full-cue data and evaluated on held-out trials with 75\% cue dropout. All 20 independently generated simulation networks met the full-cue SNR criterion for every mechanism. Ten were excluded because at least one associative-recurrence response at 75\% dropout did not reach the fixed-point tolerance within 500 iterations; the failure rate was one to three of 160 responses in each excluded network, totaling 14 of 1,600 responses. No failure occurred at lower dropout or in either one-pass feedforward branch. The remaining 10 networks were eligible for every complete paired held-out contrast; these networks are simulation replicates, not biological observations. Among eligible networks, mean calibration SNRs were 2.001, 2.002, and 2.001 for associative recurrence, feedforward denoising, and divisive normalization, respectively.

Relative to feedforward denoising, associative recurrence increased completion cosine ($\Delta=0.036$, $p=0.0009766$, $q=0.007812$), nearest-target accuracy ($\Delta=0.016$, $p=0.005859$, $q=0.01562$), split-half reliability ($\Delta=0.064$, $p=0.003906$, $q=0.01562$), and target-geometry similarity ($\Delta=0.053$, $p=0.02148$, $q=0.04297$; each $n=10$ networks). The comparison with divisive normalization was not significant for completion cosine ($\Delta=-0.033$, $p=0.9756$, $q=0.9756$), nearest-target accuracy ($\Delta=-1.11\times10^{-17}$, $p=0.6172$, $q=0.7054$), split-half reliability ($\Delta=0.036$, $p=0.05176$, $q=0.07552$), or target-geometry similarity ($\Delta=0.041$, $p=0.05664$, $q=0.07552$). Because joint eligibility was determined after held-out convergence checks, these comparisons are conditional on the retained networks. Thus, the recurrent model exceeded one matched alternative but not the normalization control, limiting the result to a qualified computational comparison rather than evidence for a specific biological mechanism.

\section{Materials and Methods}
% ============================================================
\label{sec:methods}
\subsection{Datasets and preprocessing}

\textit{Steinmetz recordings.}
We analyzed the three Neuropixels archives from \cite{steinmetz2019distributed}. The loader required all 39 archived sessions to be present. Sessions with fewer than 60 trials and area--session recordings with fewer than 10 units were excluded. ``Root,'' missing, and empty anatomical labels were omitted; Allen-area acronyms were assigned to predefined coarse regions, with unlisted acronyms assigned to ``Other.'' Missing mouse identifiers were assigned session-specific labels rather than pooled, preventing unknown animals from being treated as the same mouse but not recovering unrecorded repeated-animal structure.

For each area--session, responses were summed over 0--500~ms after stimulus onset. Bin width and stimulus onset were read from the archive; missing values used the documented 10-ms bin width and onset bin 50, respectively. NaN counts were replaced by zero. Conditions were the observed ordered pairs of left and right visual contrast. The confirmatory Steinmetz analyses used these raw 500-ms spike counts: trial vectors were neither square-root transformed nor L2-normalized. Cosine distances normalized vector direction internally, but count magnitude remained available for behavioral coupling and quality-control measures. Mean firing rate was the mean count divided by window duration. Fano factor was the mean within-condition count variance-to-mean ratio over condition--unit cells with positive means. Signal variance was across-condition variance of condition means after subtracting mean-estimation variance and clipping at zero; noise variance was pooled within-condition variance, and SNR was their ratio.

\textit{Bolding--Franks recordings.}
We used the PCX-1 Simul and TeLC-PCX ExperimentCatalog files and processed \texttt{.efd} recordings from \cite{Bolding2018}. Catalog \texttt{FT} and \texttt{LT} values selected the inclusive awake-series range. Responses were taken from the first post-odor sniff. \texttt{MultiCycleSpikeRate} was used when present and \texttt{MultiCycleSpikeCount} otherwise; response type was read from the file and never inferred from scale. The multi-unit channel was excluded by retaining Python unit index 1 onward (MATLAB unit 2 onward). Both banks from acquisition 170621 were excluded under the documented paired hardware-failure rule. No additional low-rate recording exclusion or median-rate threshold was implemented.

\subsection{Geometric reliability and sampling calibration}

For two independent condition-mean matrices, $\mathbf{M}^{(1)}$ and $\mathbf{M}^{(2)}$, we computed upper-triangular cosine-distance vectors and defined geometric reliability as
\begin{equation}
\widehat{\mathcal S}
=
\rho_{\mathrm{Spearman}}\!\left[
 \operatorname{pdist}_{\cos}(\mathbf{M}^{(1)}),
 \operatorname{pdist}_{\cos}(\mathbf{M}^{(2)})
\right].
\end{equation}
At least three retained conditions and non-degenerate finite RDMs were required.

The sampling-calibrated Steinmetz analysis jointly matched unit number and repetitions per condition. The primary specification used 10 units and two repetitions per condition per half; grids of 10, 20, or 30 units and one, two, or three repetitions per half were retained as sensitivity analyses. For each area--session and grid cell, 100 deterministic Monte Carlo draws sampled units without replacement and randomly allocated exactly matched condition-specific repetitions to two halves. Conditions lacking the required repetitions were omitted. The area--session estimate was the mean reliability across valid draws; draw SDs and quantiles describe Monte Carlo variation, not biological confidence intervals.

A stationary residual-bootstrap benchmark preserved each condition mean and empirical neuron--neuron residual dependence. Within each condition, complete residual vectors were sampled independently with replacement and assigned without retaining trial order. The procedure therefore preserves the condition-specific empirical joint distribution of population responses in expectation but removes serial dependence, slow gain changes, and other trial-order structure. Adding a resampled residual to its condition mean reconstructs an observed nonnegative count vector exactly, apart from floating-point roundoff, so the zero clipping step does not alter the raw-count bootstrap distribution. Matched split-half reliability was recomputed in 100 simulations, and the expected value under this fitted exchangeable-residual model and the observed-minus-expected difference were recorded. This is a condition-wise residual bootstrap, not a temporal block bootstrap. Removing serial structure can raise expected reliability when real trial-order nonstationarity makes sampled halves differ, but the direction and size of the effect are model-dependent and do not identify a temporal mechanism.

As a conventional distance-strength benchmark, crossnobis distances were computed on separately sampled units. For each eligible condition, two trials reserved exclusively for covariance estimation supplied one residual difference divided by $\sqrt{2}$; the next two matched sets supplied independent condition means. A single Ledoit--Wolf precision estimate was fitted to the residual rows and applied to every condition pair:
\begin{equation}
d_{ij}
=
\frac{(\bar{\mathbf{x}}^{(1)}_i-\bar{\mathbf{x}}^{(1)}_j)^\mathsf{T}
\widehat{\mathbf{\Sigma}}^{-1}
(\bar{\mathbf{x}}^{(2)}_i-\bar{\mathbf{x}}^{(2)}_j)}
{p},
\end{equation}
where $p$ is the sampled unit count. Crossnobis is a crossvalidated distance, not a second reliability coefficient.

\subsection{Regional inference}

Primary regional tests first averaged area--session outcomes and quality covariates within each mouse and coarse region. Outcomes and covariates were rank-standardized, mouse fixed blocks were included, and the regional partial-$F$ statistic was tested with 9,999 restricted-model residual permutations among regions within each mouse. Confirmatory regional endpoints were matched reliability adjusted for SNR, mean rate, and Fano factor, and excess reliability over the stationary benchmark. Unadjusted matched reliability and quality-adjusted mean crossnobis distance were secondary benchmarks.

Regional descriptive intervals used 10,000 hierarchical bootstrap samples, resampling mice and then sessions within mouse. Same-session coarse-region contrasts first averaged fine areas within session and region, formed paired regional differences, and averaged multiple session contrasts within mouse before a two-sided sign-flip test. Exact signs were enumerated for at most 16 mice; otherwise 100,000 random sign patterns were used. Leave-one-session-out and leave-one-mouse-out estimates were sensitivity analyses. Trials, units, conditions, and Monte Carlo draws were never biological replicates.

\subsection{Outer cross-fitting, behavioral coupling, and decoders}

Within each Steinmetz session, repetitions of every contrast-pair condition were randomly allocated to four folds without reference to neural activity. Only a multiple of four repetitions was retained, giving equal support in every fold. The designated primary partition estimated reliability from folds 0 and 1 and all behavioral and decoder outcomes from disjoint folds 2 and 3. The reciprocal direction and four additional deterministic partitions were reported separately as sensitivity analyses.

Each area--session used 10 random draws of 10 units; metrics were averaged over draws before inference. Choice coupling was the Spearman correlation between response direction and the L2 norm of the raw count vector on finite, non-zero-response trials, after centering magnitude within stimulus condition. At least 10 responded trials and both response classes were required.

Stimulus and response-direction decoders were class-balanced logistic regressions with the \texttt{lbfgs} solver and 500 maximum iterations. Standardization was fitted only on the training fold. Balanced accuracy was computed for train-fold-2/test-fold-3 and the reverse direction and averaged over the directions that produced finite estimates; the response decoder excluded no-response trials. Within an otherwise valid direction, test classes absent from training were retained in the test labels and therefore reduced balanced accuracy rather than being discarded.

The primary association between reliability and each outcome was a rank regression with fine-area fixed effects. Standard errors were clustered by mouse, and two-sided $p$-values used restricted-null mouse-level wild sign randomization: all signs were enumerated for at most 16 mice and 99,999 signs were sampled otherwise. Adjusted models additionally included ranked SNR, mean rate, Fano factor, and recorded unit count. Area--sessions were regression observations, but mouse clusters---not unit draws or partitions---determined randomization inference.

\subsection{Within-session temporal geometry}

Temporal analyses used raw counts, 10 sampled units, two repetitions per condition per half, and 100 unit draws. For each condition, the two earliest and two latest repetitions formed a common four-trial pool. The blocked estimate compared the early and late condition means. The interleaved estimate divided the same early and late trials between both halves, so each half contained one early and one late repetition. The area--session temporal decrement was interleaved minus blocked reliability, averaged over unit draws.

For the lag analysis, repetitions of each condition were divided in recording order into four successive blocks using approximately equal array splits. An RDM was formed for each block, and RDM correlations were computed for every block pair sharing at least three conditions. Unit draws were averaged within area--session and block pair before inference. Similarity was then averaged by mouse and block lag, and a linear slope across available lags was estimated for each mouse.

The primary interleaved-minus-blocked contrast was tested on mouse-mean differences with a one-sided sign-flip test; lag decline was tested on mouse-specific slopes in the negative direction. A secondary quality-adjusted regional test used the restricted within-mouse permutation procedure above. Trial order indexed time: these analyses assess within-session nonstationarity and do not measure elapsed-time effects or multi-day representational drift.

\subsection{Bolding concentration generalization}

Only loading A was analyzed because it provided a documented two-odor, four-concentration panel. Zero-based valves 1--4 encoded ethyl butyrate at 0.03, 0.10, 0.30, and 1.00\% v/v, and valves 9--12 encoded hexanal at the same concentrations. Recordings were required to contain at least six trials in each of the eight conditions and at least six units that were finite on every selected trial, non-negative, and had strictly positive mean response.

Responses were square-root transformed before unit matching. Complete same-session pairs comprised exactly one OB and one PCx recording in Simul, or one TeLC and one control PCx recording in TeLC-PCX. Pairs failed if both catalog animal identifiers were available but disagreed. Each pair was matched to the smaller unit count. For each member, 200 unit subsets were sampled without replacement; each trial vector was L2-normalized only after unit selection, and metrics were averaged over draws.

Split-half reliability used alternating repetitions within each odor-by-concentration condition. Odor-axis consistency was the mean of the six pairwise cosine similarities among the ethyl-butyrate-minus-hexanal mean-response axes at the four concentrations. The concentration-distance ratio was the mean cosine distance between concentrations of the same odor divided by the mean distance between different odors. Leave-one-concentration-out decoding trained a class-balanced, train-standardized logistic regression on three concentrations and tested odor identity on the fourth; balanced accuracy was averaged only when all four held-out concentrations were valid.

Benefits were defined as PCx minus OB or control minus TeLC for reliability, axis consistency, and decoding, and in the reverse direction for the distance ratio, for which lower values were favorable. One-sided paired sign-flip tests used animal-averaged contrasts only when every pair had matching catalog animal identifiers. If animal identity was unavailable, session-pair tests were explicitly descriptive because independence across sessions could not be established. Split-half reliability was diagnostic; the three concentration-generalization metrics were confirmatory only when animal-level independence was available.

\subsection{Competing mechanism models}

We compared feedforward subspace denoising, divisive normalization, and associative recurrence in independently generated effective rate networks. Script defaults were 20 networks, eight orthonormal input conditions, 50 input channels, and 100 output units. Network $k$ used seed $320+1009k$. Balanced bipolar target patterns defined a signed projection-rule matrix $\mathbf{Q}$, and a trained feedforward matrix mapped full cues toward those patterns. Every mechanism received the same additional feedforward nuisance component orthogonal to the stored-pattern subspace, with Frobenius norm equal to 0.5 times that of the original feedforward matrix.

For feedforward drive $\mathbf{h}$, the candidate mechanisms were
\begin{align}
\mathbf{z}_{\mathrm{FF}} &=(1-a)\mathbf{h}+a\mathbf{Qh},
& a&\in\{0.25,0.50,1.00\},\\
\mathbf{z}_{\mathrm{DN}} &=\frac{\mathbf{h}}
 {1+a\sqrt{\operatorname{mean}(\mathbf{h}^2)}},
& a&\in\{0.25,0.50,1.00,2.00\},\\
\mathbf{z}_{\mathrm{AR}} &=\mathbf{h}+a\mathbf{Q}\tanh(\mathbf{z}_{\mathrm{AR}}),
& a&\in\{0.50,1.00,1.50\}.
\label{eq:rnn}
\end{align}
The recurrent fixed point was updated with 0.8 of the previous state and 0.2 of the target state for at most 500 iterations, with convergence tolerance $10^{-8}$. Gaussian observation noise was applied after every mechanism, followed by the same $\tanh$ nonlinearity.

Parameters and noise amplitudes were selected separately for each mechanism and network using only full-cue calibration trials. Eight trials per condition were used to match a target SNR of 2.0. Noise was searched on a 17-point geometric grid from 0.01 to 4.0 and refined by bounded scalar optimization. Deterministic full-cue targets were scaled to unit RMS without consulting degraded-cue outcomes. A network was eligible only if every mechanism achieved SNR within 5\% of target without calibration convergence failures.

Held-out random streams supplied 20 trials per condition at 25, 50, and 75\% cue dropout. Masked cues were restored to the full cue's L2 norm. The same keyed uniform mask draws and standard-normal observation-noise draws were used across mechanisms, parameters, and dropout levels within a network, condition, trial, and phase; mechanism-specific calibrated noise amplitudes scaled the shared normal draws. Calibration, held-out testing, and deterministic targets used distinct phases. Networks with any held-out convergence failure were excluded as complete pairs. Outcomes were split-half cosine-RDM reliability, cosine similarity to the condition's deterministic full-cue target, nearest-target cosine-classification accuracy, and Spearman similarity between Euclidean RDMs of held-out condition means and deterministic targets.

All 20 networks passed the full-cue SNR and convergence criteria for every mechanism. The 10 exclusions occurred only during held-out testing of associative recurrence at 75\% dropout. In each excluded network, one to three of the 160 recurrent responses failed to reach the $10^{-8}$ tolerance within 500 iterations (14 failures among 1,600 responses across the excluded networks). No recurrent response failed at 25 or 50\% dropout, and neither one-pass feedforward mechanism had a held-out failure.

At 75\% dropout, associative recurrence was compared separately with each feedforward control using one-sided paired network-level sign-flip tests. The independently generated network, not the condition or trial, was the inferential unit. These signed-weight effective models do not obey Dale's law and cannot by themselves establish biological recurrence.

\subsection{Multiplicity and reproducibility}

Benjamini--Hochberg adjustment was applied within, not across, analysis families. The four regional omnibus specifications formed one family; all eligible same-session regional contrasts across matched reliability and excess-over-stationary outcomes formed another. Outer-cross-fit unadjusted and quality-adjusted tests were corrected separately across the three designated outcomes. The two temporal hypotheses formed one family. Bolding adjustment included only tests meeting both the animal-independence and confirmatory-metric criteria. The eight maximum-dropout mechanism contrasts formed one family. Sensitivity partitions, descriptive Bolding tests, hierarchical intervals, leave-one-cluster analyses, and the standalone temporal regional analysis did not create additional confirmatory claims.

The master seed was 320. Steinmetz random streams were deterministically keyed by session, area, analysis offset, and partition without Python's randomized string hash; model streams were keyed by network, condition, trial, phase, and random source. Bolding unit-subsampling draws were also deterministic under the master seed. Monte Carlo draws were integrated within their recording or network and never used to increase inferential sample size.

\section{Discussion}
\label{sec3}
Representational geometry became less reproducible as trials were separated within a session. Interleaving the same early and late trial pools increased split-half RDM similarity, and similarity declined with lag across successive trial blocks. Both effects were evaluated at the mouse level and survived correction across the two planned temporal tests. Regional differences did not survive correction, and geometric reliability did not predict choice-direction coupling or decoder performance in the designated outer cross-fit. The olfactory comparisons were descriptive. In the simulations, recurrence differed from feedforward denoising but not from divisive normalization. The evidence supports within-session nonstationarity of population geometry, but not a regional hierarchy, a behavioral association, or a unique circuit mechanism.

The temporal result extends the classical observation that repeated neural responses are variable~\citep{Schiller1976, Shadlen1998, Goris2014} from single responses to the relational structure among population responses. Its interpretation rests on the matched construction. The blocked and interleaved estimates used the same condition-specific early and late trials, the same number of repetitions, and the same sampled units; only their allocation to the two RDMs differed. A larger interleaved than blocked correlation is therefore consistent with a component of variability that evolves with trial order rather than with independent stationary noise alone. The four-block lag analysis reached the same conclusion through a complementary construction. These analyses do not identify the source of nonstationarity: slow changes in arousal, engagement, adaptation, learning, electrode stability, or other latent variables could all contribute.

Representational drift operates on a longer timescale. Studies across days or weeks show that population codes can reorganize while task information or behavior remains stable~\citep{Rule2020, Aitken2022, Qin2023, Driscoll2022}. The present data contain no longitudinal tracking of common neurons, and trial order is only an ordinal proxy for elapsed time. The observed lag dependence is therefore evidence of within-session nonstationarity, not multi-day drift. It also shows that pooling repetitions within a session can violate a stationary sampling assumption. Future studies should measure split-half RDM reliability within blocks, across blocks, and across days while tracking the same cells and behavioral state.

Sampling calibration qualifies the apparent regional ordering. At a matched 10 units and two repetitions per condition per half, raw reliability was low on average and varied descriptively among regions, with higher estimates in visual and motor areas than in hippocampal areas. However, neither the quality-adjusted regional omnibus nor the same-session pairwise comparisons survived multiplicity correction, and regional variation was not detected for reliability in excess of the stationary residual-bootstrap expectation. Thus, these data do not establish a brain-wide hierarchy of geometric reliability. The residual bootstrap is itself conditional on fitted condition means, empirical residuals, and the available trial counts. It preserves whole-trial population residuals but breaks their order and serial dependence. Because each reconstructed raw-count vector is a resampled observed vector, zero clipping is inactive except for floating-point roundoff and cannot explain the higher benchmark. The negative mean excess may reflect trial-order nonstationarity absent from the exchangeable bootstrap, but it is not a direct estimate of that process. The benchmark shows that observed split-half correlations require an explicit sampling model and matched neural population size.

The outer-cross-fitted analyses provide a stringent test of functional associations. Estimating reliability and behavioral or decoder outcomes from disjoint trials removed shared trial noise as a source of association. Under this design, neither the unadjusted nor quality-adjusted confirmatory models supported an association with choice-direction coupling, stimulus decoding, or response-direction decoding after correction. These null results are specific to this design. Splitting finite repetitions among four folds and matching to 10 units reduce precision, and the L2-magnitude choice measure captures only one coarse form of neural-behavioral coupling. The data do not support the conclusion that geometric reliability predicts behavior or outperforms decoding. Future tests should use prospectively specified behavioral readouts, more repetitions per condition, and independent data for reliability and functional outcomes.

The analyses of the Bolding and Franks data are descriptive. Simultaneous PCx and OB recordings favored PCx in the specified direction for all four concentration-generalization summaries, but only five session pairs were available. The three pairs of Control and TeLC sessions showed little reliability difference and effects opposite the predicted direction for the other summaries. Animal identifiers were unavailable, so repeated sessions could not be reduced to independent animals and the session-pair tests are not confirmatory. The present results are inconclusive about recurrence and do not alter prior evidence that piriform recurrence supports stable odor representations~\citep{Bolding2018, Bolding2020}. Identifying the effect would require a larger manipulation with known animal identities, simultaneous controls, and behavior measured across concentrations.

Circuit identification requires comparison with alternative mechanisms. Associative recurrence improved pattern completion, nearest-target accuracy, split-half reliability, and target-geometry similarity relative to feedforward subspace denoising under severe cue dropout. Yet none of those outcomes differed significantly from divisive normalization after correction. Recurrence is therefore one sufficient mechanism for stabilizing degraded representations in this model class, not a uniquely supported mechanism. The comparison is also conditional on the 10 of 20 networks that passed joint calibration and held-out convergence criteria, and the effective signed-weight networks do not obey Dale's law. Experiments that measure transient dynamics, perturbation recovery, or cell-type-specific responses could distinguish mechanisms that produce similar gains in output reliability.

Split-half RDM correlation also has broader limitations. It is a global summary and can conceal instability restricted to particular condition pairs or representational subspaces. Its value depends on the chosen distance, condition set, response window, and finite-sample allocation, even after the matching and bootstrap procedures used here. Coarse anatomical mappings can combine functionally heterogeneous fine areas, while only 10 mice determine the principal Steinmetz inference. Finally, the Steinmetz and Bolding datasets differ in task, sensory system, recording design, and available metadata, so they should not be treated as direct replications.

The study shows how an established split-data RSA reliability measure~\citep{Nili2014, Walther2016} can be compared across populations without treating unequal neuron counts, trial counts, Monte Carlo draws, or area-by-session rows as independent evidence. It also identifies replicable temporal structure in representational reliability within sessions. The null and descriptive results set practical limits: regional, behavioral, and circuit claims require stronger designs than raw split-half correlations alone.

\section*{Declarations}
\textbf{Conflict of interest} The authors declare no competing interests.

\noindent \textbf{Generative AI Usage}
Large language models (specifically the GPT, Claude, and Gemini families) were used to assist with (limited) code factoring, code review and debugging, and text reviewing, editing, and polishing. All hypotheses, experimental designs, analyses, and interpretations were independently formulated and verified by the author, and the author assumes full responsibility for all content and claims in this work.

\section*{Acknowledgments}
The author wishes to thank Padma K. and Annapoorna Raju for generously supporting the computational resources used in this work. The author extends gratitude to the many institutions and individuals whose open-source datasets, frameworks, and models were used in our work.

\section*{Code and Data Availability}
All code needed to reproduce the analyses and the computational models is publicly available at \texttt{https://github.com/prashantcraju/neuroscience-reliability}. Analyses used \texttt{shesha-geometry} version 0.2.27, available from PyPI \citep{shesha2026}. The two datasets used in the study (Steinmetz et al., 2019 and Bolding \& Franks, 2018) are both publicly available. The GitHub repository contains code for automatically downloading the Steinmetz data and instructions for downloading the Bolding and Franks data.

\bibliographystyle{APA}

\end{document}